# Embodied Approximation of the Density Classification Problem via Morphological Adaptation.

JEFF JONES[1]*

*Centre for Unconventional Computing, University of the West of England, Bristol, UK, BS16 1QY*

The Majority (or Density Classification) Problem in Cellular Automata (CA) aims to converge a string of cells to a final homogeneous state which reflects the majority of states present in the initial configuration. The problem is challenging in CA as individual cells only possess information about their own and local neighbour states. The problem is an exercise in the propagation and processing of information within a distributed computational medium. We explore whether the Majority Problem can be approximated in a similarly simple distributed computing substrate — a multi-agent model of slime mould. An initial pattern of discrete voting choices is represented by spatial arrangement of the agent population, temporarily held in-place by an attractant stimulus. When this stimulus is removed the model adapts its shape and size, moving to form a minimal distance connecting line. The final position of this line is shown — in simple examples — to successfully represent the majority vote decision, and also accurately reflects the *size* of the majority. We note properties, limitations and potential improvements to the approach before returning 'full-circle' by re-encoding this morphological adaptation approach in a simple (and more space efficient) 1D CA model.

* email: jeff.jones@uwe.ac.uk

## 1 INTRODUCTION

The majority, or density classification (DC), problem is a classic problem in the field of cellular automata (CA). The problem is summarised as follows for a 1 Dimensional CA. For $n$ number of cells, each with binary state $1$ or $0$, the aim is to evolve the CA over time (iterating the transition function for all cells with local communication radius $r$, over a number of time steps) so that the final homogeneous state of all cells ($1$ or $0$) reflects the majority of the initial configuration of cells. i.e. if more cells were initialised with $1$ than $0$, the state of all cells should converge on $1$ (and vice versa).

Gacs, Kurdyumov, and Levin first suggested [9] a binary state, $r = 3$ CA rule in which individual cells assessed the number of their neighbours (current cell $x$, $x - 1$ and $x - 2$ when cell $x$ is zero, and $x$, $x + 1$ and $x + 2$ when cell $x$ is 1), assigning the current cell $x$ the majority value. Their rule (the *GKL* rule) did not, however, guarantee an accurate appraisal of majority in all configurations. When the distribution of $0$ and $1$ cells was close to $50\%$ the rule classified approximately $70\%$ of configurations correctly. Evolutionary techniques have been used to tackle the inverse problem of discovering CA rules for the DC problem, with improvements to the GKL approach presented [18, 23]. Land and Belew subsequently reported that it is not possible to guarantee correct DC with a two-state rule [17], however, small modifications to either the rule definition (for example, the use of two rules in the approach described in [7]) or changing the specification of the output result [3] can correctly solve the problem.

Alternative approaches to DC using probabilistic CA [8, 5], CA with memory [21, 1], asynchronous CA [6], and CA with multiple states [2] have been presented. For a thorough treatment of advances and variants on the DC task, the reader is referred to [4]. Cellular Automata, however, are just one example of unconventional computing substrates where simple local interactions generate complex emergent global behaviours. Can the requisite propagation of information about local cell state density be implemented in other substrates? An embodied approach using mobile agents was attempted in [12] with a population of robotic agents (red and green, the number of each corresponding to the density distribution). Each agent attempted to avoid collision with other agents, remembering the colour of each near-miss. After five such encounters each agent changed colour to match the colour it most frequently encountered. This method had a classification error rate of 18% when the distribution was close to 0.5. In living systems it was suggested in [19] that the dynamics of plant stomata aperture adjustment corresponded to the propagation of information within density classification CA.

The dialogue between different computing substrates may itself cross-pollinate in both directions. CA have been used to model slime mould *Physarum polycephalum* [10, 22, 11] and it has recently been demonstrated that slime mould itself can approximate CA dynamics, when constrained within a suitably patterned environment [20]. This raises more general questions, including how to implement distributed computing algorithms in different substrates, and which substrates are most suitable for implementation of particular algorithms.

In an effort to explore these questions, we examine the DC task and attempt to implement it within a different distributed computing substrate. We examine how the problem must be re-encoded to be approximated in a distributed multi-agent model of slime mould which behaves as a collective material computation mechanism. We describe the approach in Section 2 and present results of the approach in Section 3, acknowledging some novel properties, limitations and scope for further improvement. The idea of cross-pollination is re-visited in Section 4 where we implement the morphological adaptation approach to the DC problem within a 1D CA which is shown to be more computationally efficient. We conclude in Section 5 by examining differences between the multi-agent morphological adaptation approach and its CA implementation, and more general comments on 'porting' methods between different computing substrates.

## 2 METHOD

We employ the multi-agent model of *P. polycephalum* introduced in [14], in which we described a large population of simple mobile agents whose behaviour was coupled via a diffusive chemoattractant lattice. Agents sense the concentration of a hypothetical 'chemical' in the lattice, orient themselves towards the locally strongest source and deposit the same chemical during forward movement. The collective movement trails spontaneously form emergent transport networks which undergo complex evolution, exhibiting minimisation and cohesion effects under a range of sensory parameter settings. The collective behaves as a virtual material which has been shown to be capable of a wide range of spatially represented unconventional computing ap-plications [15]. A full description of the model is given in the Appendix.

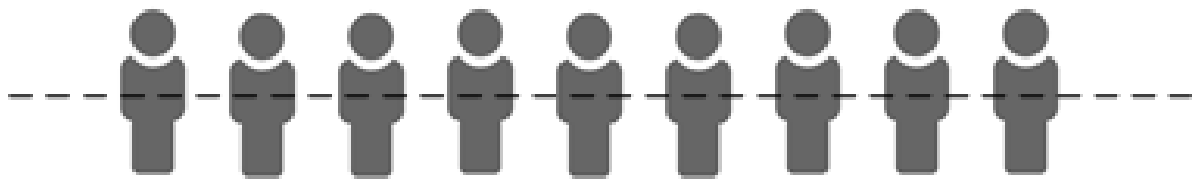

FIGURE 1
Schematic illustration of the arrangement of voters before the election.

To approximate the majority voting problem with the multi-agent approach we are required to implement the problem spatially. After the problem has been represented, the 'processing' stage is performed by the adaptation of the particle population, and the final result must also be represented by a spatial pattern. To achieve this we take a literal approach to the voting problem, i.e. we represent individual voters, and the binary choice they make (between one candidate or the other — no undecided vote is allowed) as patterns in space.

An overview of the method is given in Fig. 1. The arena is equally divided up into $n$ number of voters, where $n$ is an odd number, to ensure a simple majority. Each voter is spaced equally in a horizontal pattern across the arena. Before the vote takes place all voters are vertically placed along the $y$ position (dotted line) at half the arena height. To vote, each of the voters must take one step forwards or backwards. After the voters have selected their preferred candidate the vertical position of each voter indicates their selection Fig. 2. A high vertical position indicates one choice, in this case the candidate "Clanton". A low vertical position indicates the alternate choice, in this case the candidate "Tramp". In a conventional arithmetically calculated poll, the winner can be ascertained by simply counting the number of votes for each candidate and the candidate with the majority of votes wins the election. In Fig 2 the winner is candidate Clanton with 6 votes to 3.

The pattern of voting is represented in the agent population by a band of the virtual material which passes through all voters, connecting each voter with its immediate neighbour (Fig. 3a). Periodic boundary conditions are enforced to connect the first voter with the last. The pattern of the band can be considered as a square wave, where wave frequency is related to the number of consecutive Clanton or Tramp voters. This voter pattern is then used as the initial configuration pattern for the multi-agent population (Fig. 3b). A population of 3000 particles is initialised at random positions along the configuration pattern (only one agent may occupy a single lattice site). At the beginning of each experiment the material is initially held in its original configuration for 20 steps by projecting attractant in the initial band configuration pattern into the diffusive lattice. The attraction of the agent particles composing the band to this stimuli prevents adaptation of the band. Note that there are some small gaps in the original agent pattern shown in Fig. 3b. These gaps in the 'material' are closed as the population initially grows in size in response to the attractant stimulus. Because of the periodic boundary conditions, the population behaves as a single band of material.

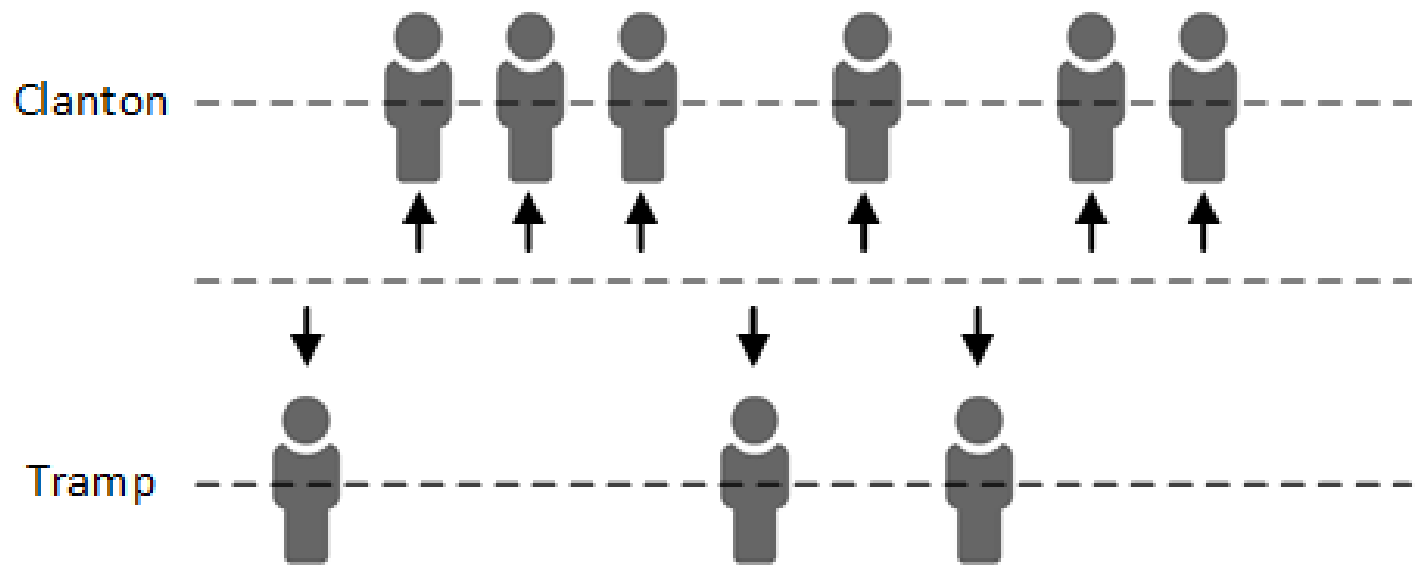

FIGURE 2
Schematic illustration of how voters select either candidate on election day by moving forwards or backwards. No undecided votes are allowed.

## 3 RESULTS

An example experimental run is shown in Fig. 4. After the attractant stimulus is removed we can see that the multi-agent population self-organises to form a network path, a curved line representing the band of material. Individual particles move along this pattern depositing trail into the lattice as they move. Because the shape of the pattern is initially tortuous, any particles incidentally taking a shorter path along the band will have this route amplified because the shorter path tends to accrue more trail than a longer path (agent particles all move at an identical speed). This positive feedback process continues and the population collectively adapts its shape and population size as the material relaxes from the initial profile. Adaptation occurs more quickly at changes in the shape profile (for example, the corners in Fig. 4a) and is less prevalent

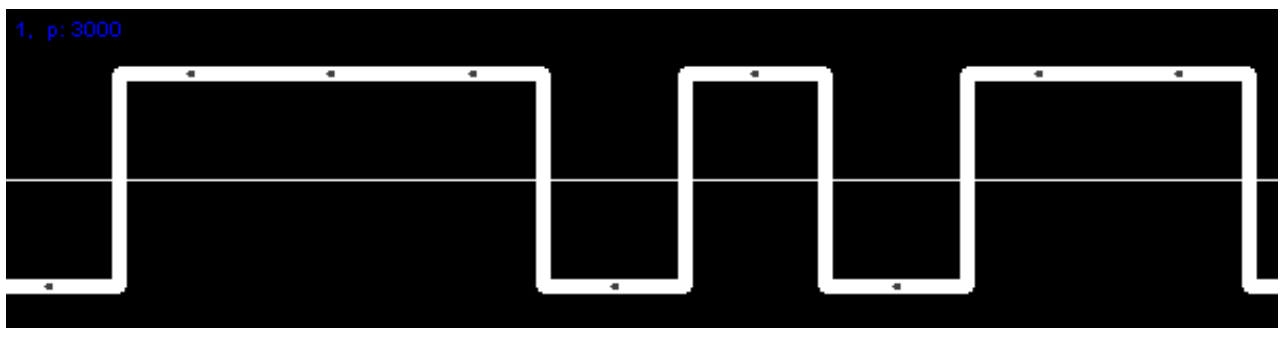

(a) voter pattern

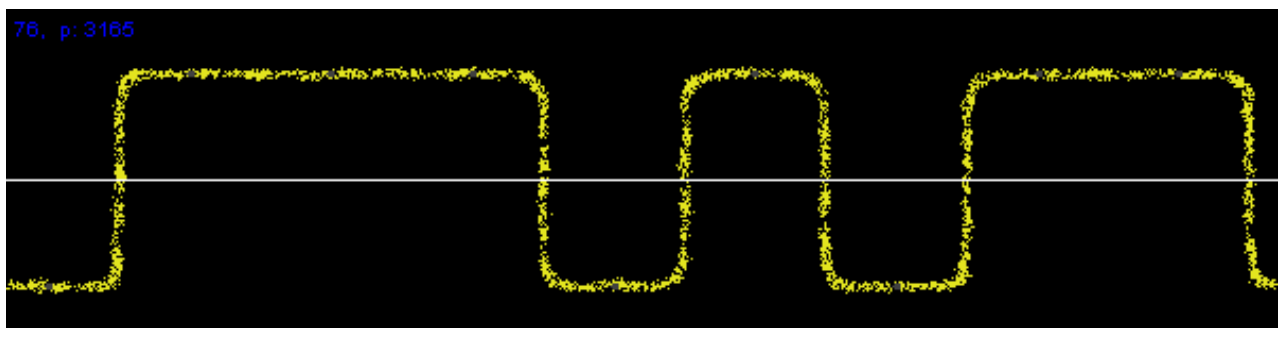

(b) initialisation pattern

FIGURE 3

How the majority classification problem is encoded in the multi-agent approach. a) voter positions (dark dots) corresponding to the vote in Fig. 2 are either above the horizontal line (candidate Clanton) or below the horizontal line (candidate Tramp). Voters are linked by straight connecting lines (white lines), b) agent population is initialised along the white line path and held in place by attractant projection for 20 scheduler steps.

along long, unchanging lines. Extremal vertical regions (sharp peaks and troughs above and below the centre line) relax over time and move towards the centre position.

Initially the adaptation is local, with rapid interaction between neighbouring voters (4b-d). When these sharp discontinuities are removed, however, the interactions between more distant voters indicated by the low frequency peaks and troughs occurs more slowly ((4)e). When the relaxation is complete, the population approximates a straight line. The position of this 'agent' line (given by the mean $y$ position of all agents) indicates which candidate has a majority in the election (4f). If the agent line is above the horizontal line then candidate Clanton has won. If the agent line is below the horizontal line then candidate Tramp has won.

A plot of various measurements recorded during the experiment (at every 50 scheduler steps) is shown in Fig. 5. The first plot shows the population size over 50,000 scheduler steps (Fig. 5a). The plot increases initially as the gaps in the initial pattern are filled in, and the material band thickens in response to the initial pattern stimulus. The population size then rapidly decreases as local interactions between individual voters occur (at locations of relatively high amplitude and high frequency) and the material shrinks in size. The population shrinkage slows as both the frequency and amplitude of discontinuities in the band decreases until the population size stabilises. The stable population size corresponds to a straightening of the material band.

The plot in Fig. 5b shows the evolution of the range of thickness of the band, i.e. the range in height between the uppermost part of the band, across the entire arena, and the lowermost part. As the band relaxes and shortens it occupies a narrower vertical range. When the band approaches becoming a straight line the thickness is approximately 10 pixels (at the particular $SO$ value used in this example). The range of thickness of the band could therefore be used as a potential halting mechanism to stop the computation.

Fig. 5c shows the evolution of the mean vertical ($y$) position of the each member of the agent population comprising the band. This indicates the current approximate majority position, although in early stages of the experiment this is only a coarse approximation because the thickness range of the band is relatively large. Note that the majority size changes during the experiment, rising and also falling. This is due to the changing nature of the interactions within the band. Up to 6000 steps there is a steady increase in the majority as the high frequency curves are flattened (as per the shapes in Fig. 4a-d). Shortly after this time, however, there is a fall in the majority size as the long distance interactions between the remaining peaks and troughs occur (Fig.

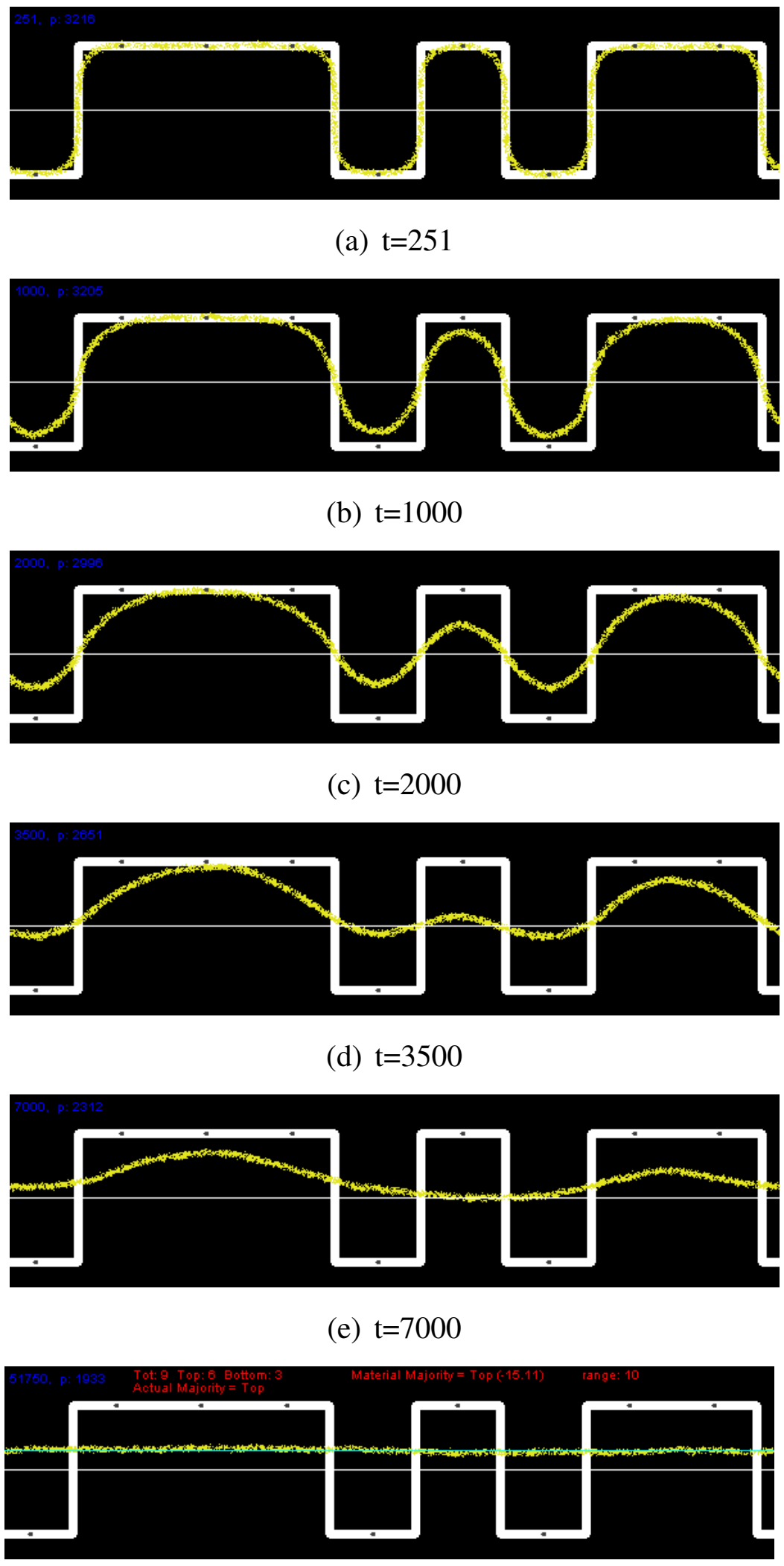

(a) t=251

(b) t=1000

(c) t=2000

(d) t=3500

(e) t=7000

(f) t=51750

FIGURE 4
Adaptation of the multi-agent population over time relaxes to solve the majority classification task. a) withdrawal of initial stimulus results in contraction of the virtual material composed of the agent population, b-f) continued relaxation, note that ‘higher frequency’ (individual) votes relax more quickly than larger clusters of identical voters, f) the final aggregate $y$ position of the agent population indicates the majority winner, depending on which side of the horizontal line the population relaxes towards.

4e-f). When the band eventually approximates a straight line, the final value of the mean vertical position of all particles indicates which side of the central line (i.e. which candidate has the majority). Note also that the distance of the majority line from the central horizontal line also approximates the *size* of the majority. The accuracy of this majority decision and majority size approximation will be assessed in more detail later.

How reliable is the morphological adaptation approach in its approximation of the majority problem? Table 1 shows results obtained during 30 simulated elections where 9 voters had to choose between the two candidates each time. The actual votes were cast randomly, resulting in 30 different pattern configuration profiles for each election. These configurations were used as the pattern to initialise the multi-agent population comprising the virtual material and adaptation of the collective was initiated. Each experiment was automatically halted when the thickness range of the material band was $\leq 10$. The results show that the morphological adaptation mechanism accurately calculated the winner of the 'election' in all 30 experiments. The mean run time for each experiment was approximately 58000 scheduler steps and the mean final population size was 1945 particles.

Not only did the mechanism correctly estimate the winner, but the final vertical position of the band also closely matched the actual winning majority in each election. The mean error for all experiments was 1.33% (maximum error 4.53%, minimum error 0.02%, standard deviation 1.33). The correlation between the winning election majority and estimated majority based on the final vertical band position was 0.986 (Pearson Product-Moment Correlation Coefficient).

### 3.1 Limitations and Potential Improvements

It should be noted that the approximation of the majority problem by this approach is not particularly efficient, in terms of time taken or space availability. The length of time to fully relax the material to a degree where it can be interpreted as a straight line is significant (approximately 58000 scheduler steps, for 9 voter experiments) and the majority of this time is spent on long range interactions between the voters. When the number of voters was increased to 19 the majority was again correct each time. However, the mean computation 'time' (over 10 runs) increased to 175000 scheduler steps (mean error for all experiments was 1.42% (maximum error 5.00%, minimum error 0.27%, standard deviation 1.37). This large increase in computation time is due to the increased distance required to propagate information along the material band.

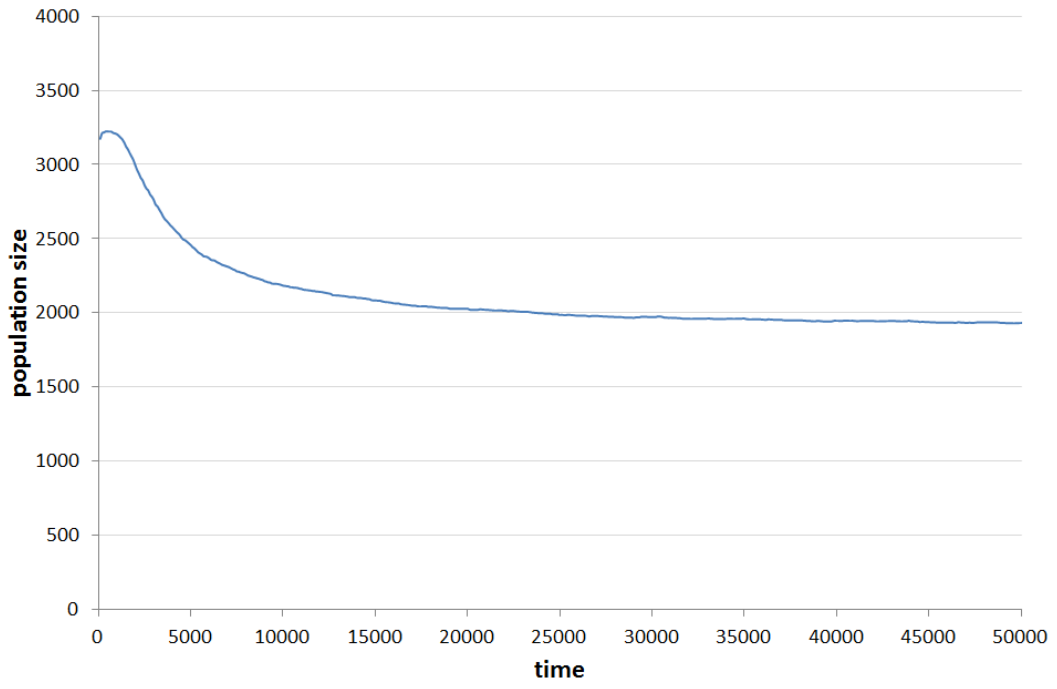

(a) population

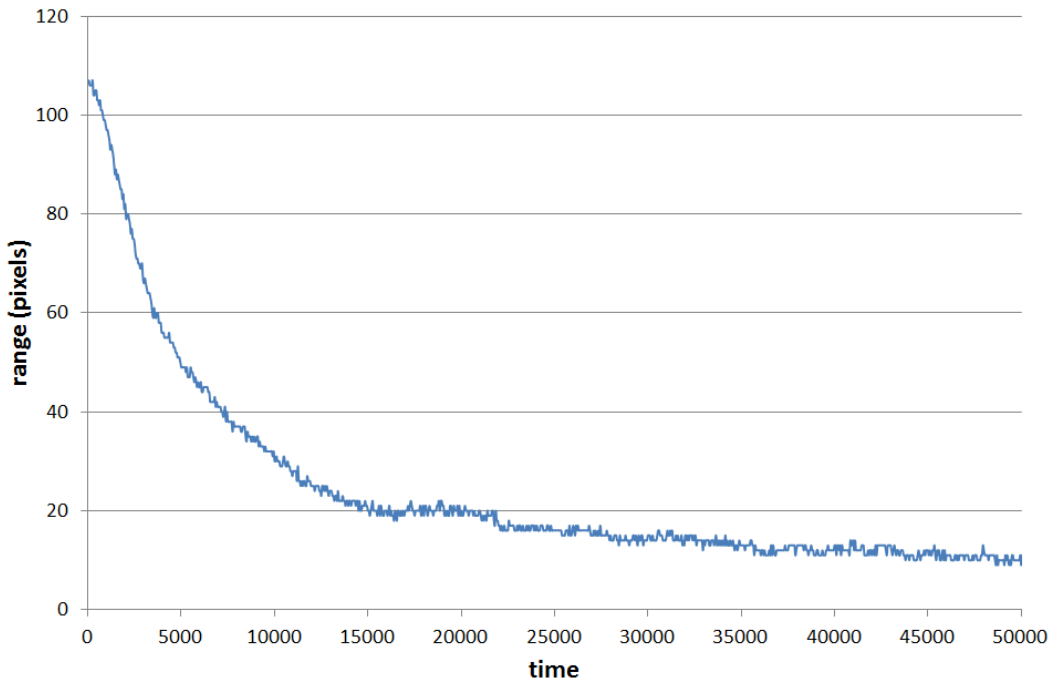

(b) thickness range

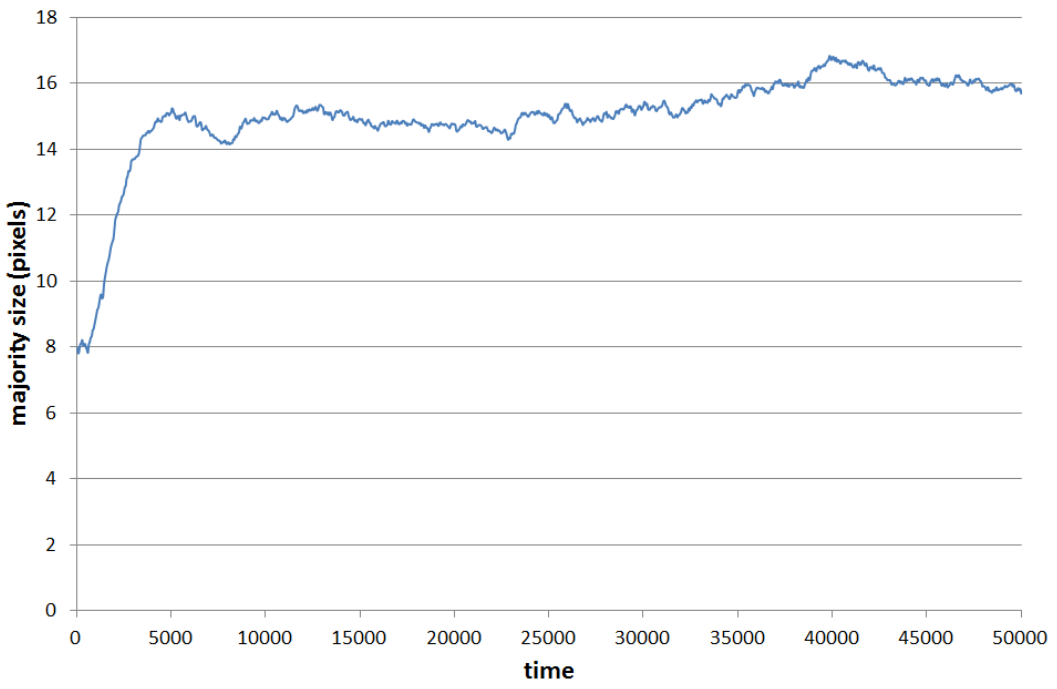

(c) majority size

FIGURE 5
Measurements sampled during evolution of Fig. 4. a) population decreases as material band shortens, b) the thickness range of the material band reduces over time, c) the mean particle $y$ position indicates the majority size (see text for details).

| | Election | | | | | | Multi-agent Morphological Adaptation | | |
|---|---|---|---|---|---|---|---|---|---|
| Experiment | Tramp Votes | Tramp % | Clanton Votes | Clanton % | Winner | Majority Size (%) | Winner | Majority Size (%) | Error (%) |
| 1 | 1 | 11.11 | 8 | 88.89 | Clanton | 88.89 | Clanton | 85.19 | 3.70 |
| 2 | 4 | 44.44 | 5 | 55.56 | Clanton | 55.56 | Clanton | 56.37 | 0.81 |
| 3 | 3 | 33.33 | 6 | 66.67 | Clanton | 66.67 | Clanton | 62.4 | 4.27 |
| 4 | 1 | 11.11 | 8 | 88.89 | Clanton | 88.89 | Clanton | 88.43 | 0.46 |
| 5 | 4 | 44.44 | 5 | 55.56 | Clanton | 55.56 | Clanton | 56.53 | 0.97 |
| 6 | 5 | 55.56 | 4 | 44.44 | Tramp | 55.56 | Tramp | 56 | 0.44 |
| 7 | 7 | 77.78 | 2 | 22.22 | Tramp | 77.78 | Tramp | 76.88 | 0.90 |
| 8 | 7 | 77.78 | 2 | 22.22 | Tramp | 77.78 | Tramp | 78.45 | 0.67 |
| 9 | 3 | 33.33 | 6 | 66.67 | Clanton | 66.67 | Clanton | 66.64 | 0.03 |
| 10 | 4 | 44.44 | 5 | 55.56 | Clanton | 55.56 | Clanton | 54.52 | 1.04 |
| 11 | 5 | 55.56 | 4 | 44.44 | Tramp | 55.56 | Tramp | 55.96 | 0.40 |
| 12 | 6 | 66.67 | 3 | 33.33 | Tramp | 66.67 | Tramp | 68.13 | 1.46 |
| 13 | 2 | 22.22 | 7 | 77.78 | Clanton | 77.78 | Clanton | 77.83 | 0.05 |
| 14 | 4 | 44.44 | 5 | 55.56 | Clanton | 55.56 | Clanton | 55.1 | 0.46 |
| 15 | 3 | 33.33 | 6 | 66.67 | Clanton | 66.67 | Clanton | 62.14 | 4.53 |
| 16 | 4 | 44.44 | 5 | 55.56 | Clanton | 55.56 | Clanton | 52.6 | 2.96 |
| 17 | 8 | 88.89 | 1 | 11.11 | Tramp | 88.89 | Tramp | 88.92 | 0.03 |
| 18 | 5 | 55.56 | 4 | 44.44 | Tramp | 55.56 | Tramp | 57.03 | 1.47 |
| 19 | 5 | 55.56 | 4 | 44.44 | Tramp | 55.56 | Tramp | 57.43 | 1.87 |
| 20 | 5 | 55.56 | 4 | 44.44 | Tramp | 55.56 | Tramp | 58.84 | 3.28 |
| 21 | 7 | 77.78 | 2 | 22.22 | Tramp | 77.78 | Tramp | 78.94 | 1.16 |
| 22 | 5 | 55.56 | 4 | 44.44 | Tramp | 55.56 | Tramp | 55.9 | 0.34 |
| 23 | 4 | 44.44 | 5 | 55.56 | Clanton | 55.56 | Clanton | 55.71 | 0.15 |
| 24 | 4 | 44.44 | 5 | 55.56 | Clanton | 55.56 | Clanton | 55.84 | 0.28 |
| 25 | 4 | 44.44 | 5 | 55.56 | Clanton | 55.56 | Clanton | 55.89 | 0.33 |
| 26 | 6 | 66.67 | 3 | 33.33 | Tramp | 66.67 | Tramp | 67.97 | 1.30 |
| 27 | 4 | 44.44 | 5 | 55.56 | Clanton | 55.56 | Clanton | 53.53 | 2.03 |
| 28 | 6 | 66.67 | 3 | 33.33 | Tramp | 66.67 | Tramp | 67.9 | 1.23 |
| 29 | 6 | 66.67 | 3 | 33.33 | Tramp | 66.67 | Tramp | 69.98 | 3.31 |
| 30 | 7 | 77.78 | 2 | 22.22 | Tramp | 77.78 | Tramp | 77.76 | 0.02 |

TABLE 1

Results of 30 experiments with 30 random elections, each with 9 voters, using the multi-agent adaptation approach. Left side of table shows results of ‘real’ election (number of votes and distribution of voters selected randomly). Right side of table shows estimation of majority by the multi-agent morphological adaptation approach. In every case the adaptation approximation resulted in the correct winner of the election and a good estimate of majority size.

The performance of the adaptation approach could potentially be improved in a number of ways. By increasing the value of the $SO$ parameter we can see faster adaptation of the material. However this is at the expense of a thicker band of material, thus necessitating a larger thickness range threshold, and a more coarse approximation (Fig. 6a). Increasing the thickness of the band reduces the time of computation to approximately 10000 scheduler steps for 9 voter experiments. It is also possible to increase the speed of adaptation by reducing voter size, i.e. give each voter representation a smaller amount of space (Fig. 6b). Adaptation in a small arena occurs more quickly than in a large arena due to the reduced interaction distance. There is a limit to this optimisation, however. For large $SO$ values combined with small voter sizes there is the potential for interaction between separate parts of the band due to the increased sensor scale. This causes the band to temporarily fuse at nearby vertical regions, contaminating the end result Fig. 6c and d.

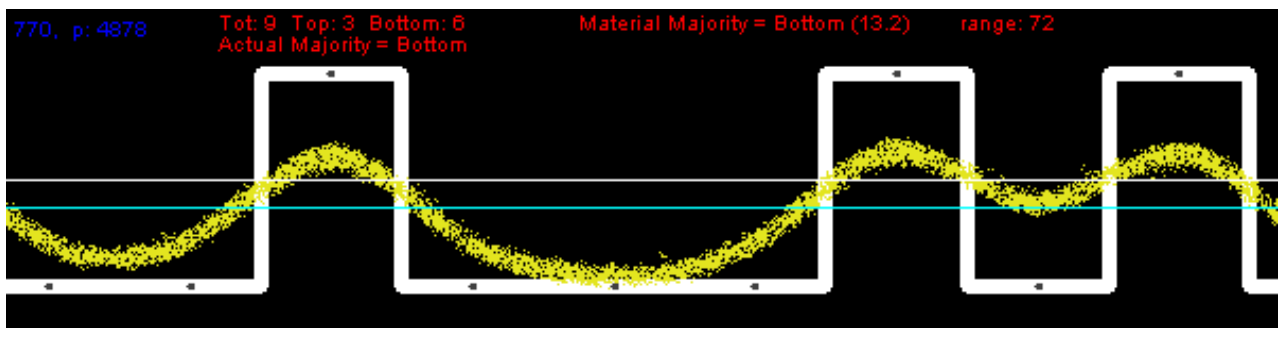

(a) 9 voters $SO$ 15, arena width 600 pixels

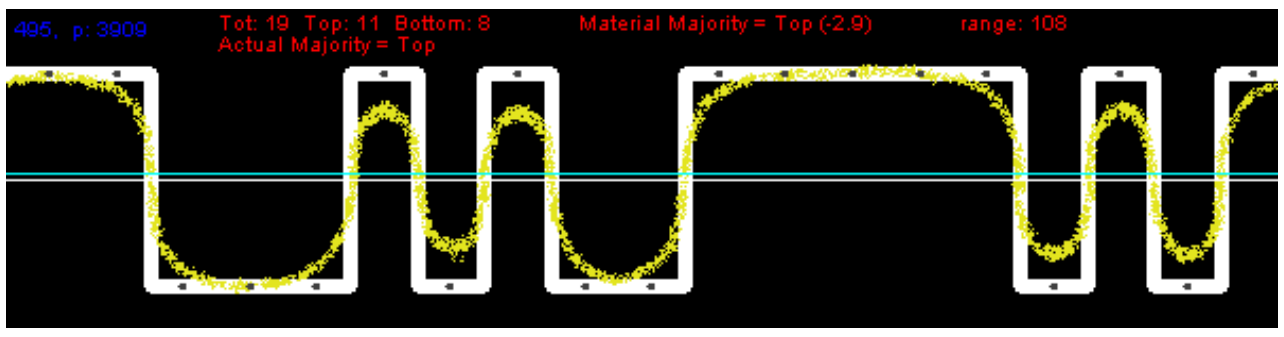

(b) 19 voters $SO$ 5, arena width 600 pixels

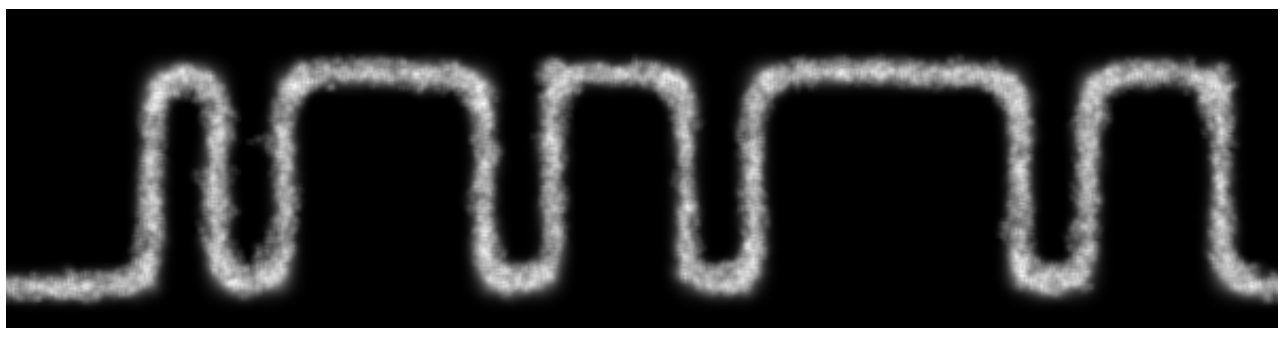

(c) 19 voters $SO$15t=41, arena width 600 pixels

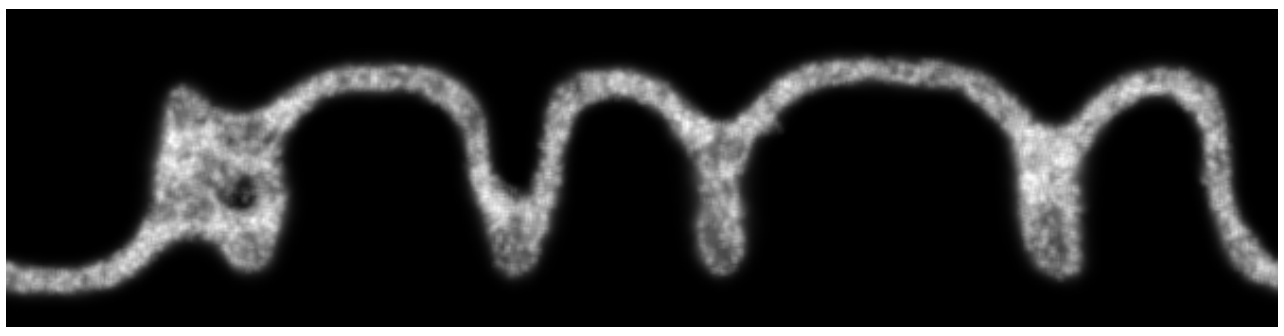

(d) 19 voters $SO$15 t=138, arena width 600 pixels

FIGURE 6
Potential methods of improving performance of the morphological adaptation method. a) increasing the $SO$ parameter to 15 results in a thicker band of material and faster adaptation, b) reducing the space requirement for a single voter allows more voters (19 in this case) to fit in a small arena space, c-d) High $SO$ scale and small voter size results in adhesion and fusion of nearby regions of material.

## 4 MIMICKING MORPHOLOGICAL ADAPTATION IN CELLULAR AUTOMATA

The original 1D CA implementations of the majority problem utilised binary states and propagation of information using a variety of rules, most common the so-called majority rules where a simple majority of the local states (typically the current cell and nearest neighbours). Can the morphological adaptation approach as used in the multi-agent system itself be implemented in CA?

We implemented a 1D, radius 1 CA with periodic boundary conditions and an odd number of cells where each cell value was randomly set to either 0 or 100. The cell value indicated a vote for either one of two candidates and was stored with double precision floating point values. Evolution of the CA was implemented synchronously as follows. The next cell state was the simple arithmetic mean of the current cell and the immediate ($r = 1$) neighbouring cells. The evolution of the CA proceeded as shown in Fig. 7 where the cell value is represented as greyscale intensity, showing a smearing out of the extremal high and low intensity values as the evolution continues (note that the 0-100 values were re-mapped to the range of 0-255 for visualisation purposes). Eventually the entire CA field stabilises on a certain greyscale value and the computation halts when the thickness range of the band is $<$ $0.01$. A plot of intensity across all cells (from a different experiment) is shown in Fig. 8, which shows an approximation of the morphological adaptation response seen in the multi-agent approach. The final stable position of the band corresponds to the majority decision. As with the multi-agent approach, the size of the actual majority corresponds closely to the final position of the 'band', i.e. its distance away from the centre line.

Because the CA implementation is more efficient than the agent approach, in terms of space and computing time, we were able to assess more runs, with larger voting populations. We performed 100 experiments, with 25, 49, 99, 113, 133, 159, 199, 265, 399 and 799 voters respectively. The correct majority decision was reached in all cases. The CA implementation of the material adaptation method is more accurate because the range size of the band is $0.01$ (whereas it was 10 in the agent approach). Also, the CA implementation is purely deterministic, whereas there is a large stochastic component to the behaviour of the individual particles in the agent approach. Mean run time of the CA implementation was strongly dependent on number of voters, as indicated by the plot where $r = 1$ in Fig. 9,a. In the multi-agent approach, the 'thickness' of the band could be increased to reduce the computation time.

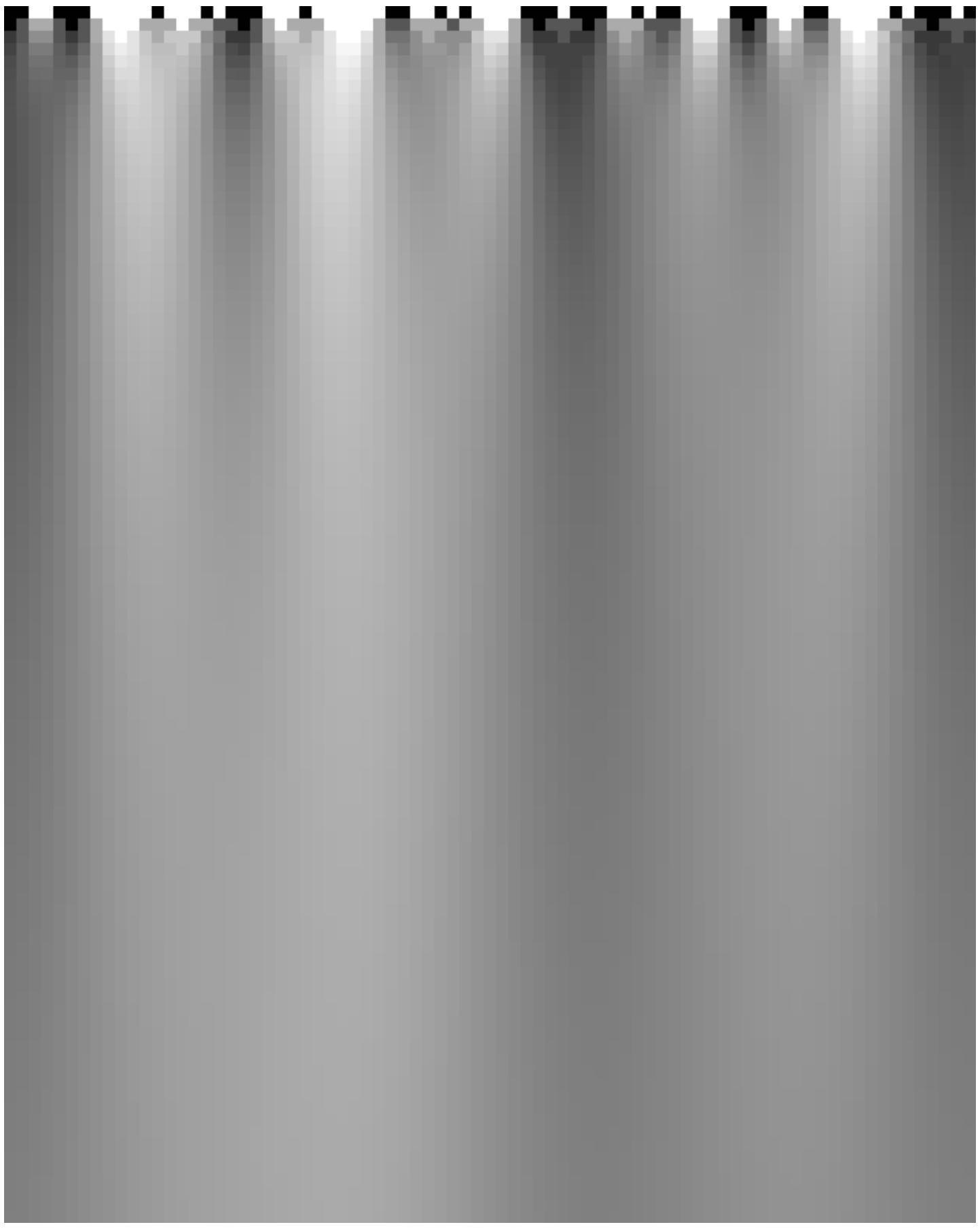

FIGURE 7
Mimicking morphological adaptation approach using a simple cellular automaton. Evolution of a 79 cell CA with periodic boundary conditions, time proceeds downwards. Simple averaging transition rule results in a smoothing of initial discrete values. Mean intensity of brightness across all cells corresponds to ‘band position’ in the multi-agent approach and, its position corresponds to the majority verdict and majority size.

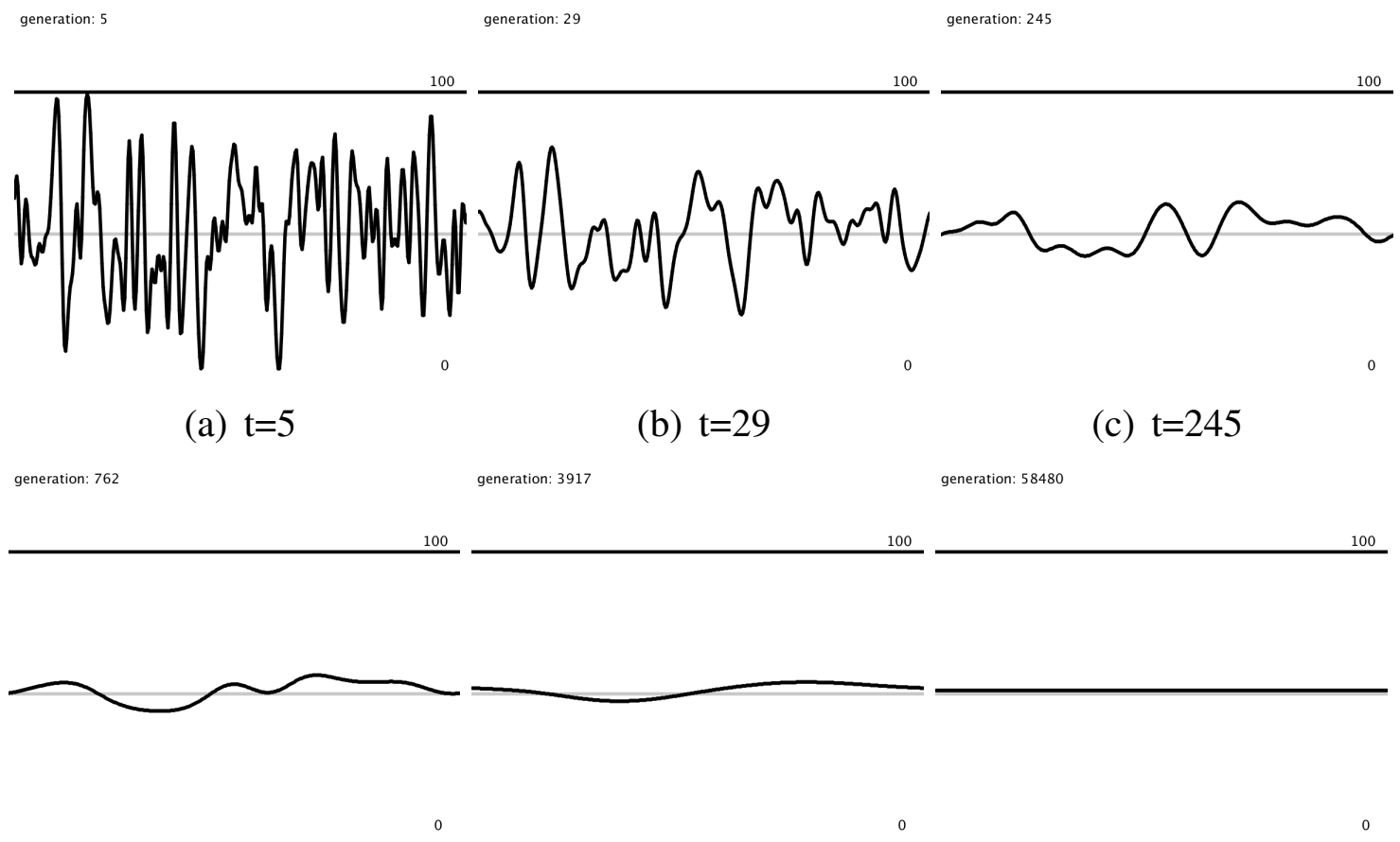

(a) t=5 (b) t=29 (c) t=245

(d) t=762 (e) t=3917 (f) t=58480

FIGURE 8
Approximation of morphological adaptation by a 1D cellular automaton 'band'. a-f) Plot of brightness intensity across 399 cells (a different example from Fig. 7) illustrates approximation of morphological adaptation and correct estimation of majority (100: 204 votes (51.13%), 0: 195 votes).

In the CA implementation, this can be approximated by increasing the radius size of the local neighbourhood. Each cell takes the arithmetic mean of every neighbour (including itself) in the neighbourhood radius $r$ centred around the current cell. Fig. 9,b illustrates how increasing the radius size reduces the mean computation time. Of course, as $r$ increases to approximate $n$, the locality is decreased until eventually the CA would simply approximate the global computation of the arithmetic mean of all cells. Nevertheless, computation time is reduced significantly for even modest values of $r$.

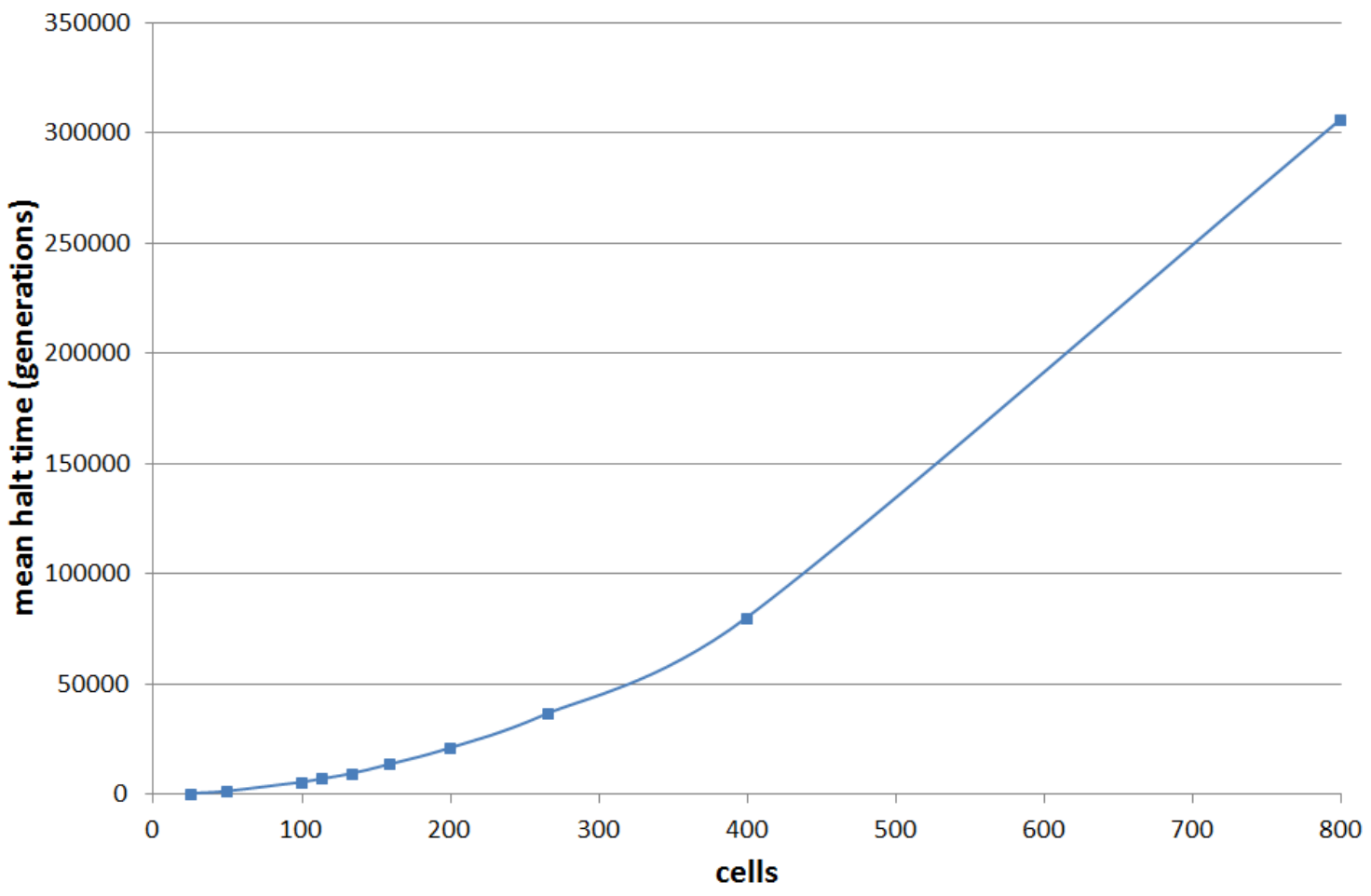

(a) halt time vs number of cells, $r = 1$

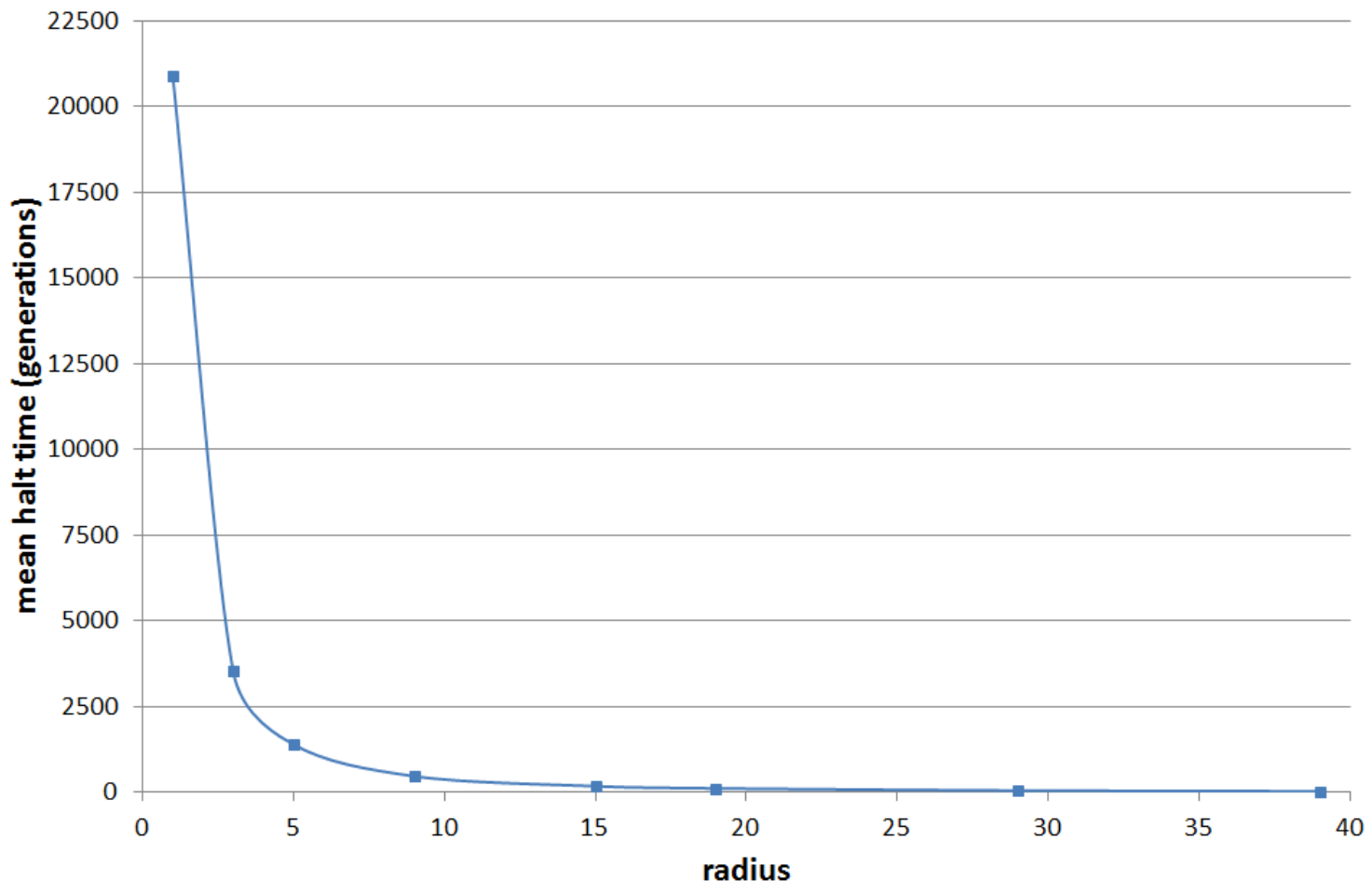

(b) halt time vs neighbourhood size, $n = 199$

FIGURE 9

Effect of cell number and neighbourhood radius on cellular automaton halt time. a) plot illustrating how mean halting time of the $r = 1$ automaton adaptation method is affected by number of cells, b) effect of neighbourhood radius on halting time, each data point is the mean halt time over 100 experiments with 199 voters, for CA with $r = 1, 3, 5, 9, 15, 19, 29, 39$.

## 5 CONCLUSIONS

The majority problem is well studied in the field of Cellular Automata, not because it is a difficult problem *per se* (in fact it is trivial to compute arithmetically in a centralised computing scheme), but because it is a good example of how problems must be encoded differently (from classical computing encoding) in unconventional computing substrates. Specifically, it requires the propagation of local information over long distances without any individual knowledge of the global state. Multi-agent systems, like CA, also exhibit the same apparent limitations of local communication and non-global knowledge but, like CA, can utilise properties of self-organisation and emergent behaviour to solve the problem. In this report we examined how the classic majority problem can be encoded in a multi-agent model of Slime mould, an organism which exhibits self-organisation and morphological adaptation from its simple component parts. By taking a literal approach to the problem we encoded the individual voting choices as contiguous spatial patterns and harnessed the adaptation phenomena, via emergent relaxation and minimisation of the virtual material. We found that — at least for simple examples — the majority decision, and indeed majority size, could be computed by this relaxation and 'read out' by the final stable position of the band.

The limitations of this encoding are the rather slow adaptation time of the material and the relatively large spatial demands necessitated by the encoding. Indeed, for larger elections the spatial requirements compound the slow relaxation time due to the increased distances involved. Although we noted improvements to the method to increase its performance (a more space efficient encoding pattern and faster adaptation with a wider material band) it should be stressed that this approach is not intended to be a 'better faster' approach, due to the inherent triviality of the arithmetic solution.

The approach does demonstrate, however, that it is possible to transfer a problem from one unconventional computing substrate (CA) to another unconventional substrate (a multi-agent system). It is interesting to note that the distributed nature of the problem is amplified in the multi-agent approach. In CA implementations, the granularity is dependent to a large extent on the number of cells in the CA (and also influenced by the neighbourhood size). In the agent approach, each 'cell' of information (initially storing each vote choice), is distributed between a much larger number of particles, all of which are moving and potentially having different neighbours in subsequent time steps. Unlike the CA approaches, where the state (whether binary or multi-state) has a discrete value and location, the agent approach *indirectly* stores the state (both the initial configuration voting pattern and the current state) within the environment as a distributed spatial pattern.

We found that the quasi-physical behaviour inherent in the multi-agent approach can also be transferred back to an analogous CA implementation which is much more space and time efficient than the multi-agent approach. In surveying the CA literature on the DC problem we find that the CA implementation of the multi-agent approach resembles the 'diffusion-amplification' approach described in [2], although the CA presented in this report has a much similar representation of diffusion and no signal amplification. We hope that the findings in this report will be of theoretical use in devising encodings for non-classical computing substrates. They may also be of potential practical use in soft-robotics systems which, because of their distributed nature, and scarce computing resources, must exploit every possible embodied computing mechanism available to them.

## 6 ACKNOWLEDGEMENTS

This research was supported by the EU research project "*Physarum* Chip: Growing Computers from Slime Mould" (FP7 ICT Ref 316366).

## 7 APPENDIX: PARTICLE MODEL DESCRIPTION

We used a multi-agent approach to generate the *Physarum*-like behaviour. This approach was chosen specifically because we wanted to reproduce the generation of complex behaviour using very simple component parts and interactions, and no special or critical component parts to generate the emergent behaviour. Although other modelling approaches, notably cellular automata, also share these properties, the direct mobile behaviour of the agent particles renders it more suitable to reproduce the flux within the plasmodium. The multi-agent particle model of *Physarum* used to generate the relaxation and adaptation [14] uses a population of coupled mobile particles with very simple behaviours, residing within a 2D diffusive lattice. The lattice stores particle positions and the concentration of a local diffusive factor referred to generically as chemoattractant. Particles deposit this chemoattractant factor when they move and also sense the local concentration of the chemoattractant during the sensory stage of the particle algorithm. Collective particle positions represent the global pattern of the material. The model runs within a multi-agent framework running on a Windows PC system. Performance is thus influenced by the speed of the PC running the framework. The particles act independently and iteration of the particle population is performed randomly to avoid any artifacts from sequential ordering.

### 7.1 Generation of Virtual Plasmodium Cohesion and Shape Adaptation

The behaviour of the particles occurs in two distinct stages, the sensory stage and the motor stage. In the sensory stage, the particles sample their local environment using three forward biased sensors whose angle from the forwards position (the sensor angle parameter, SA), and distance (sensor offset, SO) may be parametrically adjusted (Fig. 10a). The offset sensors generate local indirect coupling of sensory inputs and movement to generate the cohesion of the material. The SO distance is measured in pixels and a minimum distance of 3 pixels is required for strong local coupling to occur. For the experiments in this article we used an SO value of 5. Increasing the SO value results in a thicker band of material, a faster evolution of the adaptation and a coarser approximation of the majority line (see supplementary material for examples at increasing SO scales). During the sensory stage each particle changes its orientation to rotate (via the parameter rotation angle, RA) towards the strongest local source of chemoattractant (Fig. 10b). Variations in both SA and RA parameters have been shown to generate a wide range of reaction-diffusion patterns [13] and for these experiments we used SA 90 and RA 45 which results in stronger and more rapid adaptation of the virtual material. After the sensory stage, each particle executes the motor stage and attempts to move forwards in its current orientation (an angle from 0–360 degrees) by a single pixel forwards. Each lattice site may only store a single particle and particles deposit chemoattractant into the lattice (5 units per step) only in the event of a successful forwards movement. If the next chosen site is already occupied, the move is abandoned and the particle selects a new random direction.

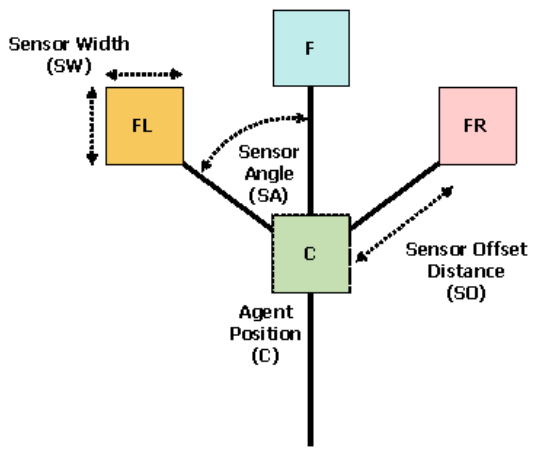

(a)

```
[Sensory stage]

- Sample chemoattractant map values
- if (F > FL) && (F > FR)
        - Continue facing same direction
- Else if (F < FL) && (F < FR)
        Rotate by RA towards larger of FL and FR
- Else if (FL < FR)
        Rotate right by RA
- Else if (FR < FL)
        Rotate left by RA
- Else
        Continue facing same direction
```

(b)

FIGURE 10
Architecture of a single particle of the virtual material and its sensory algorithm. (a) Morphology showing agent position ‘C’ and offset sensor positions (FL, F, FR), (b) Algorithm for particle sensory stage.

### 7.2 Problem Data Representation

For the spatial representation of the majority problem, the voter choices were represented by a polyline connecting the voters in a square wave pattern. The arena was divided into equal spacing $s$ relating to the number of voters (arena width, $w/x$, where $x$ is the number of voters). Each voter was connected in the following way. A horizontal line centred on each voter of width $s$ passed through each voter. The edge of each voter was connected with the next voter. If the next voter selected the opposing candidate the voters were connected by a vertical line, otherwise the horizontal line continued. Periodic boundary conditions connected the first and last voter by the same method. This resulted in a polyline representing the initial stimulus pattern. This pattern was used to maintain the population in place by projecting attractant into the lattice at a value of 2.55 units per polyline pixel. This projection prevented adaptation of the population. Agent particle movement was halted for a short time (20 scheduler steps) to allow attraction to the initial data stimuli and growth of the population to fill any gaps in the band pattern. To initiate complete relaxation of the material the stimulus pattern was removed from the lattice.

### 7.3 Material Shrinkage Mechanism

Relaxation and adaptation of the virtual material was implemented via tests executed at regular intervals as follows. If there were 1 to 10 particles in a $9 \times 9$ neighbourhood of a particle, and the particle had moved forwards successfully, the particle attempted to divide into two if there was a space available at a randomly selected empty location in the immediate $3 \times 3$ neighbourhood surrounding the particle. For shrinkage: If there were 0 to 24 particles in a $5 \times 5$ neighbourhood of a particle the particle survived, otherwise it was deleted. Deletion of a particle left a vacant space at that location which was filled by nearby particle movement, thus causing the band of material to shrink slightly. As the process continued the material shrank and adapted its morphology to shorten the band.

The frequency at which the growth and shrinkage of the population is executed determines a turnover rate for the particles. The frequency of testing for particle division and particle removal was every 2 scheduler steps. This relatively high frequency (compared to other applications using the virtual material approach, e.g. [16]) is due to the strong shrinkage invoked by the particular $SA$ and $RA$ combination used, necessitating a high adaptation frequency to maintain connectivity of the band of material as it adapted its shape.

### 7.4 Halting Mechanism

To decide when to halt the adaptation and read the result, we assessed the range of thickness of the band pattern. Specifically we refer to the difference between the maximum and minimum $y$ coordinate of all particles comprising the band. This range was compared to the thickness of a straight (non-meandering) band pattern (approximately 10 pixels thick at $SO$ 5). The adaptation was halted when band thickness was $\leq$ 10 pixels and the final mean position of the band was recorded and compared to the actual voting pattern to calculate the majority decision and majority size estimates.